\documentclass[twocolumn,10pt,a4paper,showpacs,preprintnumbers,amsmath,amssymb,notitlepage]{revtex4-1}
\date{March 2010}
\usepackage{epsfig}
\usepackage{latexsym}
\usepackage{amssymb}
\usepackage{booktabs}
\usepackage{graphicx}
\newcommand{\be}{\begin{equation}}
\newcommand{\ee}{\end{equation}}
\newcommand{\ba}{\begin{eqnarray}}
\newcommand{\ea}{\end{eqnarray}}
\newcommand{\bi}{\begin{itemize}}
\newcommand{\ei}{\end{itemize}}
\newcommand{\ud}{\,\mathrm{d}}

\newcommand{\half}{{\textstyle\frac{1}{2}}}

\newcommand{\<}{\langle}
\renewcommand{\>}{\rangle}
\newcommand{\eq}{Eq.~}

\newcommand{\fig}{Fig.~}

\newcommand{\la}{\label}

\newcommand{\txts}{\textstyle}

\newcommand{\bx}{\boldsymbol{x}}
\newcommand{\bp}{\boldsymbol{p}}
\newcommand{\bj}{\boldsymbol{j}}
\newcommand{\bq}{\boldsymbol{q}}
\newcommand{\bk}{\boldsymbol{k}}

\newcommand{\Epipi}{E_{\pi\pi}}

\begin{document}
\title{Lattice QCD and the Timelike Pion Form Factor}

\author{Harvey~B.~Meyer}
\affiliation{Johannes Gutenberg Universit\"at Mainz, 
    Institut f\"ur Kernphysik, 55099 Mainz, Germany}

\date{\today}

\begin{abstract}
We present a formula that allows one to calculate the pion form factor
in the timelike region $2m_\pi\leq \sqrt{s}\leq 4m_\pi$ in lattice
QCD.  The form factor quantifies the contribution of two-pion states
to the vacuum polarization. It must be known very
accurately in order to reduce the theoretical uncertainty on the anomalous
magnetic moment of the muon.  At the same time, the formula
constitutes a rare example where, in a restricted kinematic regime,
the spectral function of a conserved current can be determined from
Euclidean observables without an explicit analytic continuation.
\end{abstract}

\maketitle

\section{Introduction}

The ab initio calculation of hadron properties in lattice QCD is a
mature and ongoing research activity. For a compilation of results for
some quantities of high phenomenological impact see
Ref.~\cite{Colangelo:2010et}.  One important aspect of lattice QCD is
that it is formulated in Euclidean space. Therefore the form factors
are necessarily extracted for momentum transfers in the spacelike
region, $q^2\equiv q_0^2-\bq^2=-Q^2<0$. At first sight, it thus seems that
a form factor in the timelike region, or more generally any quantity
which involves the notion of `real time', is only accessible via
analytic continuation. The latter typically involves solving a
Fredholm equation of the first kind such as \eq(\ref{eq:DispRel}),
which represents a numerically ill-posed problem~\cite{Meyer:2011gj}.

The goal of this letter is twofold. First, to provide an example
where, in a certain kinematic regime, a spectral function (to be
defined shortly) can be extracted from the Euclidean field theory
without an explicit analytic continuation. And secondly, to lay the
ground for future calculations of the pion form factor in the timelike
region, which may impact the theoretical determination of the
anomalous magnetic moment of the muon.  Given the high likelihood of a
new $(g-2)_\mu$ experiment at Fermilab~\cite{E-989}, this application
is particularly timely.

For the electromagnetic current 
$j^{\rm em}_\mu=\frac{2}{3}\bar u\gamma_\mu u - \frac{1}{3} \bar d \gamma_\mu d +\dots$
the spectral function is defined as 
\be
\rho_{\mu\nu}(k)\equiv \frac{1}{2\pi}
\int d^4x\, e^{ik\cdot x}
\<0| [\hat j^{\rm em}_\mu(x),\hat j^{\rm em}_\nu(0)]|0\>.
\ee
Due to current conservation and Lorentz invariance, the tensor structure
of $\rho_{\mu\nu}$ is 
\be
\rho_{\mu\nu}(k) = -(g_{\mu\nu}k^2 - k_\mu k_\nu) \cdot \rho(k^2).
\ee
The spectral density $\rho$ is non-negative. The leading hadronic contribution
to the vacuum polarization $e^2\Pi(Q^2)$ in the spacelike domain can be expressed
through it via a once-subtracted dispersion relation,
\be \la{eq:DispRel}
 \Pi(0)-\Pi(Q^2) = Q^2 \int_0^\infty \ud s \frac{\rho(s)}{s(s+Q^2)}.
\ee
The function $\Pi(Q^2)$ can be calculated in 
lattice QCD~\cite{Blum:2002ii,Aubin:2006xv,Feng:2011zk,DellaMorte:2010sw}.
On the other hand, $\rho(s)$ is related to experimental observables,
\be \la{eq:rhoR}
\rho(s) =\frac{R(s)}{12\pi^2},
\qquad
R(s) \equiv  \frac{\sigma(e^+e^-\to {\rm hadrons})}
 {4\pi \alpha(s)^2 / (3s) } .
\ee
The denominator is the treelevel cross-section $e^+e^-\to\mu^+\mu^-$
in the limit $s\gg m_\mu^2$, and we have neglected QED corrections.
At low energies, the spectral density is given by the pion form factor
defined in (\ref{eq:Fpi})~\cite{Jegerlehner:2009ry},
\be
\rho(s) = { \txts\frac{1}{48\pi^2} }
\Big(1-\frac{4m_\pi^2}{s} \Big)^{\frac{3}{2}}  |F_\pi(\sqrt{s})|^2,
\qquad |F_\pi(0)|=1.
\la{eq:RFpi}
\ee
The relation holds for $2m_\pi\leq\sqrt{s}\leq 3m_\pi$, and even up to
$4m_\pi$ if the electromagnetic current is replaced by the isospin
current in the definition of $\rho(s)$.

In infinite volume, the spectral density is a continuous function
above the two-particle threshold. In finite volume, where simulations
are carried out, it is a collection of delta functions.  What is
needed is an explicit formula that relates the energy and matrix
element of one individual state living on the 3d torus to the
infinite-volume spectral density at the same energy.  In this letter
we derive the formula
\be
|F_\pi(E)|^2 = \Big(q \phi'(q) +k \frac{\partial\delta_1(k)}{\partial k}\Big)
\frac{3\pi E^2}{2 k^5}    |A_\psi|^2 .
\la{eq:result1}
\ee
Here $E$ equals the invariant mass of the two pions, $k$ is related to $E$
via \eq(\ref{eq:Epipidef}), $\delta_{1}$ is the
scattering phase shift in the unit isospin, $p$-wave channel and $A_\psi$ is a
vector-current matrix element between the vacuum and a two-pion state
$|\psi^a_\sigma\>$ of energy $E$ on the torus, see
\eq(\ref{eq:Apsi}). Finally, $q\equiv \frac{kL}{2\pi}$ and $\phi$ is a
known kinematic function~\cite{Luscher:1991cf}.  The scattering phase
$\delta_{1}(k)$ can be extracted (see~\cite{Feng:2010es,Aoki:2007rd} 
and Refs. therein) from the finite-volume spectrum using
the L\"uscher formula~\cite{Luscher:1990ux,Luscher:1991cf},
\eq(\ref{eq:Luscher_cond}).

The derivation of \eq(\ref{eq:result1}) is closely related to work by
Lellouch-L\"uscher~\cite{Lellouch:2000pv} on the matrix element
determining the $K\to\pi\pi$ decay rate.
\eq(\ref{eq:RFpi}--\ref{eq:result1}) show that this formula,
remarkably, allows one to extract a spectral function from Euclidean
field theory without an explicit analytic continuation.

\section{Preliminaries}

We will assume here that isospin is an exact symmetry of QCD, 
and focus on the vector isovector channel
\be \la{eq:symchann}
I=1, \qquad J^{PC}=1^{--}.
\ee
Both in experiment and in
the Euclidean theory, the symmetry of the final state can be
selected. Therefore, in spite of the physical photon not having
definite isospin quantum numbers, we will consider the
case of a gauge boson coupling to the isospin current
\be
j^a_\mu = \bar \psi \gamma_\mu {\txts\frac{\tau^a}{2}} \psi,
\qquad \psi= \Big(\begin{array}{c} u \\ d \end{array}\Big).
\ee
In infinite volume, and with quark masses set to their physical
values, the states in the symmetry channel (\ref{eq:symchann}) are
necessarily two-pion states for center-of-mass energies
\be \la{eq:Erange}
2m_\pi \leq  E  \leq 4m_\pi.
\ee
We are thus led to consider pion scattering in this energy range.  
First, the normalization of single-pion states is 
\be \la{eq:ooNorm}
\< k\, a| k'\, b\> = \delta_{ab}(2\pi)^3 \,2E_k\delta(\bk-\bk').
\ee
The $T_I$ matrix describing the scattering of two pions is defined in
full generality in (\cite{Luscher:1991cf}, Sec. 2.1). 
It carries an isospin index $I=0,1,2$;
here we are interested in the $I=1$ channel.
The $T_1$ matrix can be expanded in partial waves with amplitudes $t_{I\ell}$,
with corresponding phase shifts $\delta_{I\ell}$.
We restrict ourselves to the $\ell=1$ partial wave in the elastic regime,
\ba \la{eq:T1}
T_1 &=&  48\pi  \Epipi \,  t_{11}\,\cos\theta\,,
\qquad
t_{11} = \frac{e^{2i\delta_{11}}-1}{2ik_\pi}. 
\ea
From now on the phase shift will simply be written $\delta_1$.
When the pions couple to the photon, they can be produced for example
in the reaction $e^+e^-\to\gamma^*\to \pi\pi$.  
With $|(\pi_{\bp}\pi_{\bp'})^a\> = 
\frac{\epsilon^{abc}}{\sqrt{2}}  |\pi^b_{\bp} \pi^c_{\bp'}\>$,
the pion form factor in the timelike region is defined
as~\cite{OConnell:1995wf}
\ba \la{eq:Fpi}
\!\!\!\!
\<0| \hat\bj^a | (\pi_{\bp}\pi_{\bp'})^b,{\rm in}\> 
&=& -\<(\pi_{\bp}\pi_{\bp'})^b,{\rm out}|\hat\bj^a| 0\> 
\\
&=& \delta^{ab} \sqrt{2} i  \, (\bp' - \bp)\, F_\pi(\Epipi).
\nonumber
\ea

The two-pion states  on a three-dimensional torus of dimensions
$L\times L\times L$ at vanishing total momentum have been studied in
detail in~\cite{Luscher:1991cf}.  The vector states $(\ell=1)$ 
are found exclusively in the $T_1$ irreducible representation of the cubic
group. Their norm will be taken to be unity.
We denote by $E_{\pi\pi}$ the energy of one such state. 
The effective momentum of the pion $k_\pi$ is then defined by the
equation
\be \la{eq:Epipidef}
\Epipi = 2\sqrt{m_\pi^2 + k_\pi^2}.
\ee

The discrete values of $k_\pi\doteq k$ in the box and the infinite-volume phase
shifts are related by~\cite{Luscher:1990ux,Luscher:1991cf}
\ba \la{eq:Luscher_cond}
\delta_1(k) + \phi(q) = n\pi, 
\qquad \qquad 
q\equiv \frac{kL}{2\pi}.
\ea
The function $\phi(q)$ is defined by $\tan\phi(q) =
-\frac{\pi^{{3}/{2}}q}{{\cal Z}_{00}(1;q^2)}$, where ${\cal
  Z}_{00}(1;q^2)$ is the analytic continuation in $s$ of ${\cal
  Z}_{00}(s;q^2) = \frac{1}{\sqrt{4\pi}} \sum_{\boldsymbol
  n\in\mathbb{Z}^3} \frac{1}{(\boldsymbol{n}^2-q^2)^s}$.

\section{QCD coupled to $SU(2)_{I}$ gauge bosons in the broken phase}

The theory we consider is QCD with at least two degenerate light
flavors of quarks $u$ and $d$.  The other flavors are assumed to be
sufficiently massive that the only unit-isospin states in the interval
(\ref{eq:Erange}) are the two-pion states. This is realized for
physical values of the quark masses.

We now wish to establish \eq(\ref{eq:result1}) in the regime where the
gaps between finite-volume energy eigenstates are substantial, which
is the situation that can realistically be achieved in Monte-Carlo
simulations. The idea is to couple the quarks infinitesimally (the
coupling will be denoted by $e$) to an external vector particle with
mass in the range (\ref{eq:Erange}). On one hand, when it is
degenerate with a two-pion energy eigenstate in the box, a level
splitting of order $e$ occurs; on the other hand, in infinite volume
the resonant production of the vector particle in pion scattering
leads to an O($e$) change in the phase shift. The phase shift and the
energy levels in the box must be in correspondence through the
L\"uscher formula (\ref{eq:Luscher_cond}) both before and after
switching on the coupling $e$. \eq(\ref{eq:result1}) follows from the
difference of these relations. The proof is thus almost identical to
the proof of the Lellouch-L\"uscher formula for the kaon
decay~\cite{Lellouch:2000pv}. The only qualitative differences are
that we are in a different symmetry channel and, more importantly,
that we invoke an external particle.  A similar formula for a matrix
element involving a two-particle state was derived
in~\cite{Lin:2001ek}, in the large-volume regime where individual
states are too narrowly spaced to be individually resolved.

A concrete model for the massive vector boson is not required in the
proof, but it may be reassuring for the validity of the argument that
the situation sketched above can be realized within a renormalizable
field theory.  A specific realization, then, involves gauging the
isospin symmetry (this was, incidentally, the original motivation of
Yang and Mills~\cite{Yang:1954ek} to introduce non-Abelian gauge
theories).  There are then three gauge bosons of the group
$SU(2)_I$. Since we want them to be massive, we assume that the gauge
group is spontaneously broken by a Higgs mechanism  
at a scale well above $\Lambda_{\rm QCD}$, the scalar
field being in the fundamental representation of  SU(2)$_I$. The three
gauge bosons $W^a$ then form a degenerate triplet of mass $M$, which
we assume to be in the range (\ref{eq:Erange}). They have the same
quantum numbers as the $\rho$ meson and couple to the isospin current.

\subsection{Two-pion energy levels in finite volume}
We start with the system in finite volume and study the effect of the
massive gauge bosons on the low-lying spectrum, which consists of two-pion states.
The gauge boson field operator reads 
\be
A^b_\mu(\bx) = 
\sum_{\bk}\sum_{\sigma=1}^3 
\frac{e_\mu^\sigma(\bk)}{\sqrt{2E_kL^3}} 
(a^b_{\bk,\sigma} e^{i\bk\cdot\bx} +a^{b\dagger}_{\bk,\sigma} e^{-i\bk\cdot\bx} ),
\ee
with $[a^a_{\bk,\sigma},a^{b\dagger}_{\bk',\sigma'}] =
\delta^{ab} \delta_{\sigma\sigma'} \delta_{\bk\bk'}$ and 
the polarization vectors $e_\mu^\sigma$ are such that at $\bk=0$,
$e_0^\sigma=0$ and $e_j^\sigma = \delta_{\sigma j}$.  
The coupling to the massive gauge bosons is treated as a perturbation, with 
\be \la{eq:pertn}
H_{\rm int} = + e \int j^{a\mu}(\bx) A^a_\mu(\bx).
\ee
The box size $L$ is now chosen such
that the energy eigenvalue of a unit-norm, non-degenerate state $|\psi^a_\sigma\>$ of the two-pion system 
overall at rest in the
box satisfies $E_{\pi\pi}=M$ when $e$ is set to zero. Upon switching on the SU(2)$_{I}$ gauge 
interaction, we apply degenerate perturbation theory of
quantum mechanics. The two-pion state mixes maximally with the massive
gauge boson state $|W^a_\sigma\>\equiv a^{a\dagger}_{\boldsymbol{0},\sigma} |0\>$, 
since they are initially degenerate, and the splitting between
the two resulting energy levels is given by the off-diagonal matrix element, 
\be \la{eq:Epipi}
E_{\pi\pi}^{\pm} = M \pm |{\cal A}| 
\ee 
with 
\ba \la{eq:Q}
 \< \psi^a_\sigma| H_{\rm int} |W^b_{\sigma'}\> 
=\delta^{ab}\,\delta_{\sigma\sigma'} {\cal A} 
&=&
\frac{-e}{\sqrt{2M}} \,A_\psi  \,\delta^{ab}\,\delta_{\sigma\sigma'},
\\
\la{eq:Apsi}
{L^{3/2}}  \<\psi_\sigma^a| \hat j_{\sigma'}^b(\bx) |0\>
&=& \delta^{ab} \delta_{\sigma\sigma'} A_\psi.
\ea

\subsection{Pion-pion scattering in infinite volume}

The coupling of quarks to the massive gauge bosons affects their scattering amplitude.
At ordinary energies, the effect is O($e^2$). However, at the energy (\ref{eq:Epipi})
the effect is enhanced to O($e$), due to the resonant production of a massive gauge boson,
whose propagator reads
\be
\frac{i\delta^{ab}}{p^2-M^2+i\epsilon}\, \left(\frac{p_\mu p_\nu}{M^2} - g_{\mu\nu}\right).
\ee
The change of the scattering amplitude due to this process then reads, 
at the energy (\ref{eq:Epipi}) and with center-of-mass frame kinematics,
\ba 
&& \Delta T_1\Big((\pi_{\bp}\pi_{-\bp})^a\to{W}{\to}(\pi_{\bp'}\pi_{-\bp'})^b\Big) 
\\ &&
= 
\<(\pi_{\bp'}\pi_{-\bp'})^b,{\rm out}|\bj^c| 0\>
\cdot 
\frac{e^2\delta^{cd}}{q^2-M^2}
\cdot
\<0| \bj^d | (\pi_{\bp}\pi_{-\bp})^a,{\rm in}\>,
\nonumber
\ea
where $q^2 - M^2 = \pm  2M|{\cal A}| + {\rm O}({\cal A}^2)$.
Using the definition of the pion form factor in the timelike
region (\ref{eq:Fpi}) and the phase $F_\pi = |F_\pi|  e^{i\delta_1}$ due to Watson's theorem, we obtain
\be
\Delta T_1 = 
\mp \frac{e^2 \delta^{ab}}{M|{\cal A}|} |F_\pi(\Epipi=M)|^2\, 4k_\pi^2 \cos\theta\, e^{2i\delta_1}.
\ee
Next we translate this into a change in the phase shift 
via \eq(\ref{eq:T1}), 
\be \la{eq:Dd1}
\Delta\delta_{1} 
= \mp \frac{e^2}{12\pi M |{\cal A}|} |F_\pi|^2 \frac{k_\pi^3}{\Epipi}.
\ee

\subsection{Connection between the energy levels in finite-volume and  the 
scattering phase shift}

We now go back to the finite volume system where $L$ is tuned so that
$E_{\pi\pi}=M$.  Upon switching on the perturbation (\ref{eq:pertn}),
this degenerate energy level in the box splits, corresponding to $k\to
k+\Delta k$, where in view of \eq(\ref{eq:Epipidef}--\ref{eq:Epipi}),
$\Delta k =  \pm {\cal A}\, \frac{\Epipi}{4k_\pi}$.
The phase shift that corresponds to the perturbed state by 
L\"uscher's formula (\ref{eq:Luscher_cond}) differs from the phase shift 
corresponding to the unperturbed state by 
\ba
\delta_1(k) &\to &\delta_1(k) + \frac{\partial\delta_1(k)}{\partial k} \Delta k + \Delta\delta_1(k).
\ea
In \eq(\ref{eq:Luscher_cond}), this variation must be compensated by the change in the second 
term of the LHS, 
\be
\phi(q) \to \phi(q) + \phi'(q) \Delta q, \qquad \Delta q = q \frac{\Delta k}{k}.
\ee
Altogether, we then obtain the condition
\be
\Delta\delta_1(k) = -\Big(q \phi'(q) +k \frac{\partial\delta_1(k)}{\partial k}\Big) \frac{\Delta k}{k}.
\ee
Inserting expression (\ref{eq:Dd1}) for the change 
in the phase shift due to the resonant scattering, one then obtains \eq(\ref{eq:result1}).

\subsection{Consistency check in the absence of pion-pion interactions}

In the case where the two pions do not interact, the $n^{\rm th}$
two-pion energy level $\Epipi$ passes through $M$ for~\cite{Lellouch:2000pv}
\be
L =\frac{2\pi}{k_\pi} \sqrt{n}\,,\qquad 1\leq n\leq 6,
\ee
$k_\pi$ being related to $\Epipi$ by \eq(\ref{eq:Epipidef}).
From the definition of $\phi$, one sees that for $q^2\to n$ an integer,
\be
q\phi'(q)  = (2\pi)^2 \frac{q^3}{\nu_n},
\ee
where $\nu_n$ is the number of vectors $\boldsymbol{z}\in\mathbb{Z}^3$ with $\boldsymbol{z}^2=n$.
The result (\ref{eq:result1}) then becomes for non-interacting pions
\be
|F_\pi(\Epipi=M)|^2 = \frac{3}{4\nu_n} \frac{M^2L^3}{k_\pi^2} |A_\psi|^2.
\la{eq:result_free}
\ee

One can check this last equation directly. Take for instance as a
two-pion state on the torus $|\psi_\sigma^a\>$ given by 
\ba \la{eq:psifree}
|\psi^a_3\> &=& \frac{\epsilon^{abc}}{\sqrt{2}} 
a^{b\dagger}_{\bp} a^{c\dagger}_{-\bp} |0\>,
\qquad
\left[a_{\bp}^{a}  ,a_{\bp'}^{b\dagger}\right]   = \delta^{ab}\delta_{\bp\bp'}.
\ea
with $\bp= p\hat e_3$ and $a_{\bp}^{a\dagger}$ a pion creation operator.
Thus from the difference in normalization of the finite and infinite-volume
states (\eq\ref{eq:psifree} and \ref{eq:ooNorm}) one predicts
\be
8k_\pi^2 |F_\pi|^2 = M^2 L^3 |A_\psi|^2,
\la{eq:result_free2}
\ee
which matches (\ref{eq:result_free}) for $\nu_n=6$. This is precisely 
the multiplicity $\nu_n$ of $n=1$ and $4$ for which 
the only vectors $\boldsymbol{z}$ are $(\sqrt{n},0,0)$ 
and images thereof under cubic symmetries.

\section{Illustration}

\begin{figure}[t!]
\centerline{\includegraphics*[width=0.45\textwidth]{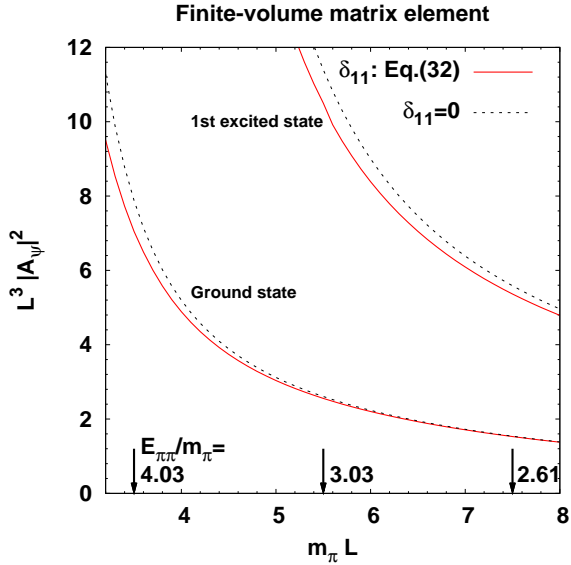}}
\vspace{-0.4cm}
\caption{\label{fig:Apsi2} Finite-volume matrix element $|A_\psi|^2$ calculated 
(a) using \eq(\ref{eq:result1})  and (b) using \eq(\ref{eq:result_free}), 
where the pion interactions are neglected.
The phenomenological input for $\delta_{1}(k)$ and $F_\pi(E)$ is given 
in \eq(\ref{eq:effrange}) and (\ref{eq:FpiExpt}).
The values of the ground state energy is indicated for three different box sizes.}
\end{figure}

Here we apply \eq(\ref{eq:result1}) to infer the dependence of the 
matrix element $A_\psi$ on $L$ from experimental scattering data.
The scattering phase $\delta_{1}(k)$ is parametrized by~\cite{Luscher:1991cf}
\ba
\frac{k^3}{E_{\pi\pi}} \cot \delta_{1}(k) &=& 
\frac{4k_\rho^5}{m_\rho^2 \Gamma_\rho} \left(1 -\frac{k^2}{k_\rho^2}  \right).
\la{eq:effrange}
\ea
with $k_\rho \equiv \half \sqrt{m_\rho^2 - 4m_\pi^2}$.
In this effective range formula, the scattering phase is thus determined 
by the mass and width of the $\rho$ meson.
Secondly, we extract the pion form factor from experimental data
compiled by the PDG~\cite{Nakamura:2010zzi}. 
At low energies the data is well described by 
\be\la{eq:FpiExpt}
|F_\pi(\Epipi)|^2 = v_0 + {v_1}\big([(\Epipi/m_\pi)^2- v_2]^2 + v_3^2\big)^{-1},
\ee
where the fit parameters for the interval $2.0m_\pi \leq \Epipi \leq 4.4m_\pi$ 
take the values $v_0 = 0.6473$, $v_1= 10.59$, $v_3 = 0.1271$,
and the $\chi^2$/d.o.f. amounts to 1.2. 
We did not fit $v_2$, but set it to the value $m_\rho/m_\pi=5.553$.

Using formula \eq(\ref{eq:result1}), we can predict the magnitude of
the finite-volume matrix element $A_\psi$ up to the inelastic
threshold in the $I=1$ channel ($4m_\pi$). The result of this exercise
is displayed in \fig(\ref{fig:Apsi2}), along with the finite-volume
matrix element one would expect if there were no interactions between
pions.  As already discussed in~\cite{Luscher:1991cf}, the size of the
matrix element $|A_\psi|^2$ is of order $L^{-3}$ in the absence of
interactions between the pions, but it would be strongly enhanced in
the vicinity of a resonance.

\section{Conclusion}

We have derived a formula which connects Euclidean observables with
the vector current spectral function in the elastic regime.  Such a
relation could have interesting applications in other systems, such as
cold Fermi gases~\cite{coldfermigas-theory}. Also, the idea to invoke
an external particle weakly coupled to QCD is likely to lead to other
important relations.  Apart from its theoretical appeal,
\eq(\ref{eq:result1}) provides a more direct way to compare lattice
calculations with experimental determinations of $\rho(s)$ than the
dispersion relation (\ref{eq:DispRel}).  To complement the low-energy
experiments constraining the leading hadronic contribution to
$(g-2)_\mu$, lattice simulations would have to reach very low pion
momenta. For this
purpose~\cite{Bedaque:2004kc,deDivitiis:2004kq,Sachrajda:2004mi}, as
well as to obtain the factor $\partial\delta_{1}(k)/\partial
k$~\cite{Kim:2010sd}, twisted boundary conditions could prove very
useful.

\noindent I thank H.~Wittig and D.~Bernecker for helpful discussions and
M.~L\"uscher for comments on the manuscript.


\bibliography{/home/meyerh/CTPHOPPER/ctphopper-home/BIBLIO/viscobib}

\end{document}